\documentclass{ws-procs975x65}

\def\ltsima{$\; \buildrel < \over \sim \;$}
\def\lsim{\lower.5ex\hbox{\ltsima}}
\def\gtsima{$\; \buildrel > \over \sim \;$}
\def\gsim{\lower.5ex\hbox{\gtsima}}
\def\msun{\mathrm{~M_{\odot}}}

\def\mdot {\dot M}
\def\xmm{{\em XMM--Newton}}

\def\hd {HD\,49798}
\def\rx {RX\,J0648.0--4418}
\def\hr {HD\,49798/RX\,J0648.0--4418}
\newcommand{\apj}{{\it ApJ}}

\newcommand{\apjs}{{\it ApJS}}
\newcommand{\apjl}{{\it ApJ}}
\newcommand{\aap}{{\it A\&A}}

\newcommand{\mnras}{{\it MNRAS}}

\newcommand{\pasj}{{\it PASJ}}

\newcommand{\apss}{{\it Ap\&SS}}

\begin{document}

\title{RX J0648.0--4418: THE FASTEST-SPINNING WHITE DWARF}

\author{S. MEREGHETTI$^*$}

\address{INAF, IASF-Milano\\
via Bassini 15, Milano I-20133, Italy\\
$^*$E-mail: sandro@iasf-milano.inaf.it
}



\begin{abstract}
\hr\ is a post common-envelope X-ray binary composed of a hot subdwarf and
one of the most massive white dwarfs with a dynamical mass measurement (1.28$\pm$0.05 $\msun$).
This white dwarf, with a spin period of 13.2 s, rotates more than twice
faster than the white dwarf in the cataclysmic variable AE Aqr.
The  current properties of these two binaries, as well as their future evolution,
are quite different, despite both contain a fast-spinning white dwarf.
\hr\ could be the  progenitor
of either a Type Ia supernova or of a non-recycled millisecond pulsars.
\end{abstract}


\bodymatter

\section{Introduction}
\label{aba:sec1}

The soft X-ray source \rx\ is optically identified with the bright star \hd\ (V=8), well known
to optical astronomers since the sixties as the brightest member of the small class of
hot subdwarf stars. \hd\ attracted much interest due to its peculiar composition:
the overabundance of nitrogen and helium,
and the low abundance of carbon and oxygen, indicate that
its  surface layers once belonged to the outer part of the
hydrogen-burning core of a massive star\cite{kud78,bis97}.
\hd\  is the stripped core of an initially much more massive and larger star that lost most of
the outer hydrogen envelope,  most likely as a consequence of non conservative mass transfer in a close binary.

Indeed, early radial velocity measurement showed that \hd\ is in a binary system with orbital
period of 1.55 days\cite{tha70}, but the nature of the companion star could not be determined due to its
faintness compared to the much brighter subdwarf.
Only in 1996, with the discovery\cite{isr97} of X-ray pulsations at 13.2 s, it became clear the the
invisible companion of \hd\ is a compact object: either a neutron star or a white dwarf.

More recently, thanks to \xmm\ X-ray timing of its regular  pulsations,
which make this system equivalent to a \emph{double-lined} spectroscopic binary,
and the discovery of the eclipse, which constrains the orbital inclination,
we could get a dynamical measure of the masses of the two stars. These are
1.50$\pm$0.05 $M_{\odot}$ for the subdwarf and
1.28$\pm$0.05 $M_{\odot}$  for its compact companion\cite{mer09}.

The high-quality \xmm\ spectra also showed that the compact object is most likely
a white dwarf\cite{mer11}. In fact, the measured X-ray bolometric luminosity of
$\sim$10$^{32}$ erg s$^{-1}$ (for the well known distance\cite{kud78} of
650 pc)  is exactly  what one would expect for a white dwarf accreting in the
stellar wind of \hd .
This hot subdwarf shows evidence of mass loss\cite{ham10} at a
rate of $\mdot_\mathrm{W}\sim3\times10^{-9} \msun$ yr$^{-1}$
and with a wind terminal velocity of   1350 km s$^{-1}$.
Despite this wind is much weaker that that of massive O type stars, its accretion onto
a neutron star would produce a luminosity much higher than the observed one.
The large pulsed fraction of $\sim$55\% rules out the possibility that the low luminosity
be due to a neutron star in the propeller stage. The X-ray spectrum of \rx\ is dominated by a very soft
blackbody component with temperature kT$\sim$40 eV and emitting radius larger than $\sim$20 km.
The size of the emitting region is too large
compared to the dimensions of a hot spot on the surface of a neutron star, required to
account for  the high pulsed fraction.
Thus \rx\ is most likely one of the most massive white dwarfs and the one with the shortest spin period.

\begin{table}
\tbl{Comparison of  \rx\ and AE Aqr}
{\begin{tabular}{@{}lcc@{}}
\toprule
     &\rx\ & AE Aqr \\
\colrule
Period             & 13.18 s                           &  33.08 s \\
Period  derivative &  $<6\times10^{-15}$ s s$^{-1}$     & 6$\times10^{-14}$ s s$^{-1}$ \\
White dwarf mass   &  1.28$\pm$0.05    $\msun$                   &  0.7--1.2  $\msun$    \\
Companion mass     &  1.50$\pm$0.05  $\msun$                      &  0.5--0.9  $\msun$  \\
Companion          &  sdO subdwarf, V=8                 & K3-5 main sequence, V$\sim$12     \\
Orbital period     &  1.55 days                       &   0.41 days   \\
Mass transfer      &    Stellar wind                     &  Roche-lobe overflow \\
X-ray luminosity   &   $\sim10^{32}$ erg s$^{-1}$   &   $\sim10^{31}$ erg s$^{-1}$    \\
Rotational energy loss rate &  $<10^{34}$ erg s$^{-1}$   &  6$\times10^{33}$ erg s$^{-1}$ \\
Distance           &  650 pc     &  100 pc    \\
\botrule
\end{tabular}}
\label{aba:tbl1}
\end{table}

\section{Comparison with AE Aqr}

It is interesting to compare the properties of \rx\ with those of another fast-spinning white dwarf:
the intermediate polar AE Aqr\cite{pat79} (see Table 1).
Cataclysmic variables of the intermediate polar class have  magnetic fields of
$\sim$5--20 MG, which
can influence the accretion flow and disrupt an eventual accretion disk, but lower than  those
found in polars, where the white dwarf rotation is synchronized with the orbital period.

AE Aqr exhibits a hard X--spectrum and strong rapid variability at all wavelengths,
contrary to \rx\ which shows a nearly constant X-ray flux, dominated  by a thermal-like component.
The spin-period of AE Aqr increases at a rate of 5.6$\times$10$^{-14}$
s s$^{-1}$, most likely as a result of the  magnetic propeller effect.
This means that the accretion stream from its companion is disrupted by the magnetic field
of the white dwarf and most of the mass is ejected \cite{wyn97}.
The fast rotation and high magnetic field of AE Aqr imply that, in principle,
particle acceleration in the white dwarf magnetosphere might occur\cite{uso93}, similar to
what happens in rotation-powered neutron stars.
The possible detection of pulsed hard X-rays (10--30 keV) in AE Aqr\cite{ter08} indicates that non-thermal
emission powered by rotational energy might be present.
The implied efficiency of $\sim$0.1\% is similar to that of radio pulsars.

While \rx\ is rotating more than twice faster than AE Aqr, there is no evidence for rotation-powered
activity. Indeed there is evidence that its magnetic field is rather small\cite{mer11} and the observed
luminosity is naturally explained by accretion from the wind of \hd .

\section{Conclusions and open questions}

The origin of the rapid rotation of the white dwarf in \rx\ is unclear.
This system is necessarily the result of evolution involving a common-envelope phase\cite{ibe85}.
If the fast rotation was not imparted at birth, significant spin-up must have occurred either
before or during the common envelope phase, since the transfer of angular momentum is
very small at the currently observed low accretion rate.
It seems more likely that spin-up occurred before the common-envelope phase,
considering its  short duration and
complicated dynamics.
A possibility is that the white dwarf was spun-up  when the expanding \hd\ was close to fill its Roche-lobe
just before the ensuing of the common-envelope phase.

The high rotational velocity has important
consequences for the future evolution of the system.
If  \rx\ hosts a CO white dwarf, it could be the progenitor of an
over-luminous type Ia supernova, since the fast rotation can
increase the mass stability limit above the value for
non-rotating stars.
Descending from relatively massive stars ($\sim$8--9 $\msun$), the delay time for such a
supernova could be relatively short, unless  the explosion is delayed by the
centrifugal effect\cite{dis11}.

If instead the  white dwarf has an ONe composition, an accretion
induced collapse might occur, leading to the formation of a
neutron star. The high spin rate and low magnetic field make
this white dwarf an ideal progenitor of a millisecond pulsar.
This could  be a promising scenario for the direct formation of  millisecond pulsars, i.e.
one not involving the recycling of old pulsars in accreting low mass X-ray binaries.

\bibliographystyle{ws-procs975x65}

\end{document}